\documentclass[twocolumn,showpacs,preprintnumbers,amsmath,amssymb]{revtex4}
\usepackage{graphicx}
\usepackage{dcolumn}
\usepackage{bm}
\begin{document}
\title{Spin-independent origin of the strongly enhanced effective
mass\\ in a dilute 2D electron system}
\author{A.~A. Shashkin$^*$, Maryam Rahimi, S. Anissimova, and S.~V.
Kravchenko}
\affiliation{Physics Department, Northeastern University, Boston,
Massachusetts 02115}
\author{V.~T. Dolgopolov}
\affiliation{Institute of Solid State Physics, Chernogolovka, Moscow
District 142432, Russia}
\author{T.~M. Klapwijk}
\affiliation{Department of Applied Physics, Delft University of
Technology, 2628 CJ Delft, The Netherlands}
\begin{abstract}
We have accurately measured the effective mass in a dilute
two-dimensional electron system in silicon by analyzing temperature
dependence of the Shubnikov-de~Haas oscillations in the
low-temperature limit. A sharp increase of the effective mass with
decreasing electron density has been observed. Using tilted magnetic
fields, we have found that the enhanced effective mass is independent
of the degree of spin polarization, which points to a
spin-independent origin of the mass enhancement and is in
contradiction with existing theories.
\end{abstract}
\maketitle

The ground state of an ideal, strongly interacting two-dimensional
(2D) electron system is predicted to be Wigner crystal \cite{chapl}.
The strength of the interactions is usually characterized by the
ratio between the Coulomb energy and the Fermi energy, $r_s=E_c/E_F$.
Assuming that the effective electron mass is equal to the band mass,
the interaction parameter $r_s$ in the single-valley case reduces to
the Wigner-Seitz radius, $1/(\pi n_s)^{1/2}a_B$ and therefore
increases as the electron density, $n_s$, decreases (here $a_B$ is
the Bohr radius in semiconductor). In the strongly-interacting limit,
no analytical theory has been developed to date. According to numeric
simulations \cite{tanatar}, Wigner crystallization is expected in a
very dilute regime, when $r_s$ reaches approximately 35. The refined
numeric simulations \cite{atta} have predicted that prior to the
crystallization, in the range of the interaction parameter
$25\lesssim r_s\lesssim35$, the ground state of the system is a
strongly correlated ferromagnetic Fermi liquid. At yet higher
electron densities, at $r_s\sim1$, the (weakly-interacting) electron
liquid is expected to be paramagnetic, with the effective mass, $m$,
and Land\'e $g$ factor renormalized by interactions. Enhancement of
$g$ and $m$ within the Fermi liquid theory is due to spin exchange
effects, with renormalization of the $g$ factor being dominant
compared to that of the effective mass \cite{renorm}. In contrast,
the dominant increase of $m$ near the onset of Wigner crystallization
follows from an alternative description of the strongly-interacting
electron system beyond the Fermi liquid approach, which also predicts
the renormalization of $m$ to be strongly sensitive to the
polarization of spins \cite{spivak,dol}.

In dilute silicon metal-oxide-semiconductor-field-effect-transistors
(MOSFETs), a strong metallic temperature dependence of the resistance
was observed a decade ago \cite{krav}. Although this anomaly was
almost immediately attributed to strong electron-electron
interactions, only after a strongly enhanced ratio of the spin and
the cyclotron splittings was found at low $n_s$ \cite{ssc} has it
become clear that the system behaves well beyond the weakly
interacting Fermi liquid. Later, it was reported that the magnetic
field required to produce complete spin polarization, $B_c\propto
n_s/gm$, tends to vanish at a finite electron density $\approx
8\times 10^{10}$~cm$^{-2}$ \cite{sh,vit}. These findings point to a
sharp increase of the spin susceptibility and possible ferromagnetic
instability in dilute silicon MOSFETs. In very dilute GaAs/AlGaAs
heterostructures, a similar behavior has been observed in both the 2D
hole \cite{gao} and electron \cite{zhu} system. Recently,
experimental results have indicated that in silicon MOSFETs it is
the {\em effective mass}, rather than the {\em $g$ factor}, that
sharply increases at low electron densities \cite{sh1}. The crucial
point for understanding these properties of dilute 2D electron
liquids is how the effective mass changes with the degree of spin
polarization.

In this Letter, we report accurate measurements of the effective mass
in a clean 2D electron system in silicon by analyzing temperature
dependence of the weak-field Shubnikov-de~Haas (SdH) oscillations in
the low-temperature limit \cite{rem3}. The effective mass is found to 
be strongly increased (by a factor of $\gtrsim3$) at low electron 
densities. Using tilted magnetic fields, we find that the value of 
the effective mass does not depend on the degree of spin 
polarization, which points to a {\em spin-independent} origin of the 
effective mass enhancement. This is in clear contradiction with 
existing theories \cite{renorm,spivak,dol}.

Measurements were made in a rotator-equipped Oxford dilution
refrigerator with a base temperature of $\approx 30$~mK on
high-mobility (100)-silicon samples similar to those previously used
in Ref.~\cite{krav00}. The resistance, $R_{xx}$, was measured by a
standard 4-terminal technique at a low frequency (0.4~Hz) to minimize
the out-of-phase signal. Excitation current was kept low enough
(0.1-0.2~nA) to ensure that measurements were taken in the linear
regime of response. Contact resistances in these samples were
minimized by using a split-gate technique that allows one to maintain
a high electron density in the vicinity of the contacts (about
$1.5\times 10^{12}$~cm$^{-2}$) regardless of its value in the main
part of the sample. Below, we show results obtained on a sample with
a peak mobility close to 3~m$^2$/Vs at $T=0.1$~K.

\begin{figure}\vspace{-0.9in}
\scalebox{0.4}{\includegraphics{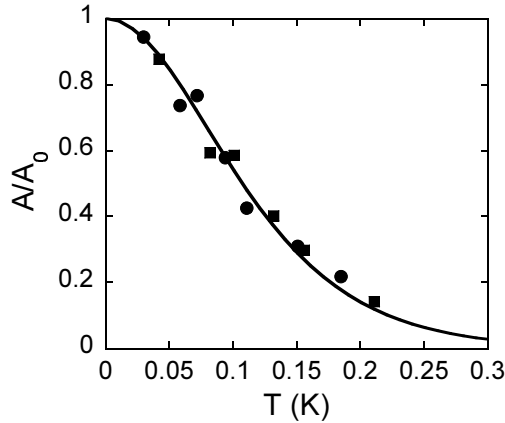}}\vspace{-1.1in}
\caption{\label{fig1} Change of the amplitude of the weak-field SdH
oscillations with temperature at $n_s=1.17\times 10^{11}$~cm$^{-2}$
for oscillation numbers $\nu=hcn_s/eB_\perp=10$ (dots) and $\nu=14$
(squares). The value of $T$ for the $\nu=10$ data is divided by the
factor of 1.4. The solid line is a fit using Eq.~(\ref{A}).}
\end{figure}

A typical temperature dependence of the amplitude, $A$, of the
weak-field (sinusoidal) SdH oscillations for the normalized
resistance, $R_{xx}/R_0$ (where $R_0$ is the average resistance), is
displayed in Fig.~\ref{fig1}. To determine the effective mass, we use 
the method of Ref.~\cite{smith} extending it to much lower electron 
densities and temperatures \cite{rem}. We fit the data for $A(T)$ 
using the formula

\begin{eqnarray}
A(T)&=&A_0\frac{2\pi^2k_BT/\hbar\omega_c}{\sinh(2\pi^2k_BT/\hbar
\omega_c)},\nonumber\\A_0&=&4\exp(-2\pi^2k_BT_D/\hbar\omega_c),
\label{A}\end{eqnarray}
where $\omega_c=eB_\perp/mc$ is the cyclotron frequency and $T_D$ is
the Dingle temperature. As the latter is related to the level width
through the expression $T_D=\hbar/2\pi k_B\tau$ (where $\tau$ is the
elastic scattering time) \cite{afs}, damping of the SdH oscillations
with temperature may be influenced by temperature-dependent $\tau$.
We have verified that in the studied low-temperature limit for
electron densities down to $\approx 1\times 10^{11}$~cm$^{-2}$,
possible corrections to the mass value caused by the temperature
dependence of $\tau$ (and hence $T_D$) are within our experimental
uncertainty which is estimated at about 10\%. Note that the amplitude
of the SdH oscillations follows the calculated curve down to the
lowest achieved temperatures, which confirms that the electrons were
in a good thermal contact with the bath and were not overheated.

\begin{figure}\vspace{-1.01in}
\scalebox{0.4}{\includegraphics{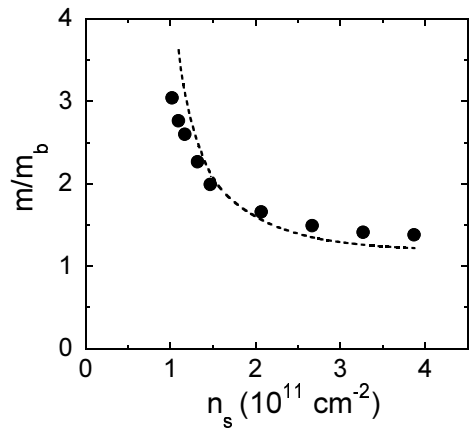}}\vspace{-1in}
\caption{\label{fig2} Dependence of the effective mass on electron
density in spin-unpolarized system. Also shown by a dashed line is
the data obtained using the method of Refs.~\cite{sh,sh1}.}
\end{figure}

In Fig.~\ref{fig2}, we show the so-determined effective mass in units
of the band mass, $m_b=0.19m_e$ (where $m_e$ is the free electron
mass), as a function of electron density. The effective mass sharply
increases with decreasing $n_s$. This is in good agreement with the
data obtained by an alternative method based on measurements of $B_c$
and of the slope of the metallic temperature dependence of
conductivity in zero magnetic field, similar to those described in
Refs.~\cite{sh,sh1}. Using that method, we also determined that our
$g$ factor is practically density-independent and is equal to
$g\approx 2.8$, which is close to $g=2$ in bulk silicon. Note that
the agreement between the results obtained using two independent
methods as well as the fact that the experimental dependence $A(T)$
follows the theoretical curve justify applicability of Eq.~(\ref{A})
to this strongly-interacting electron system.

A strong enhancement of $m$ at low $n_s$ may originate from spin
effects \cite{renorm,spivak,dol}. With the aim of probing a possible
contribution from the spin effects, we introduced a parallel magnetic
field component, $B_\parallel$, to align the electrons' spins. As the
thickness of the 2D electron system in Si MOSFETs is small compared
to the magnetic length in accessible fields, the parallel field
couples largely to the electrons' spins while the orbital effects are
suppressed \cite{sim,rem1}.

\begin{figure}\vspace{-0.93in}
\scalebox{0.4}{\includegraphics{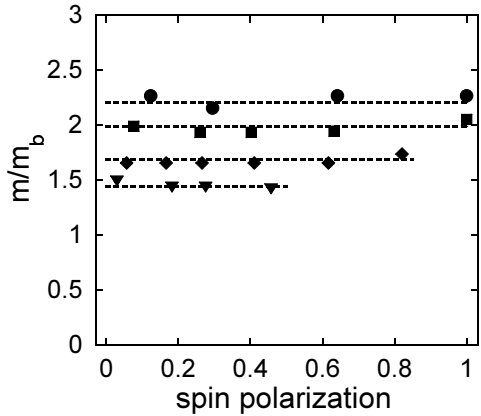}}\vspace{-1.1in}
\caption{\label{fig3} The effective mass vs the degree of spin
polarization for the following electron densities in units of
$10^{11}$~cm$^{-2}$: 1.32 (dots), 1.47 (squares), 2.07 (diamonds),
and 2.67 (triangles). The dashed lines are guides to the eye.}
\end{figure}

In Fig.~\ref{fig3}, we show the main result of the paper: the
behavior of the effective mass with the degree of spin polarization,
$p=(B_\perp^2+B_\parallel^2)^{1/2}/B_c$. Within our accuracy, the
{\em effective mass $m$ does not depend on $p$}. Therefore, the
$m(n_s)$ dependence is robust, the origin of the mass enhancement has
no relation to the electrons' spins and exchange effects \cite{rem2}.

\begin{figure}\vspace{0.07in}
\scalebox{0.4}{\includegraphics{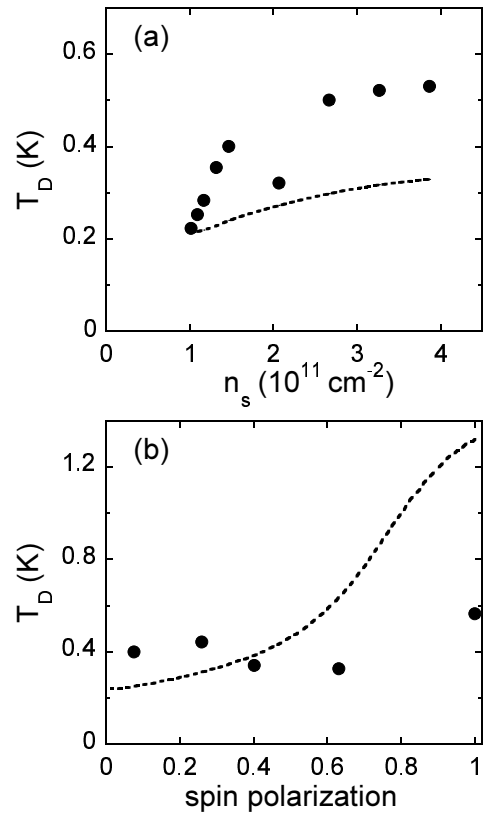}}
\caption{\label{fig4} Behavior of the Dingle temperature extracted
from SdH oscillations (dots) and that calculated from the transport
scattering time (dashed line) with electron density in
spin-unpolarized system (a) and with the degree of spin polarization
(b). In (b), the electron density is equal to $1.47\times
10^{11}$~cm$^{-2}$.}
\end{figure}

In Fig.~\ref{fig4}(a), we compare the extracted Dingle temperature,
$T_D$, with that recalculated from the electron lifetime determined
from zero-field mobility \cite{sh1}. Although the elastic scattering
time defining the Dingle temperature is in general different from the
transport scattering time, the two $T_D(n_s)$ dependences are
consistent with each other indicating dominant large-angle scattering
\cite{afs}. This indicates that the Dingle temperature decreases with
decreasing electron density, in agreement with the narrowing of the
cyclotron resonance line observed at low $n_s$ in Si MOSFETs
\cite{cr}. In contrast, Figure~\ref{fig4}(b) shows that with
increasing degree of spin polarization, the Dingle temperature
remains approximately constant, while the resistance increases by a
factor of about five. So, the straightforward correlation between the
Dingle temperature and the elastic scattering time does not hold for
fully spin-polarized electron system.

We now discuss the results obtained for the effective mass. We stress
that in the present case, the interaction parameter, $r_s$, is larger
by a factor of $2m/m_b$ than the Wigner-Seitz radius and reaches
approximately 50, which is above the theoretical estimate for the
onset of Wigner crystallization. As has already been mentioned, two
approaches to calculate the renormalization of $m$ and $g$ have been
formulated. The first one exploits the Fermi liquid model extending
it to relatively large $r_s$. Its main outcome is that the
renormalization of $g$ is large compared to that of $m$
\cite{renorm}. In the limiting case of high $r_s$, one may expect a
divergence of the $g$ factor that corresponds to the Stoner
instability. These predictions are in obvious contradiction to our
data: (i) the dilute system behavior in the regime of the strongly
enhanced susceptibility --- close to the onset of spontaneous spin
polarization and Wigner crystallization --- is governed by the 
effective mass, rather than the $g$ factor, through the interaction 
parameter $r_s$; and (ii) the insensitivity of the effective mass to 
spin effects also cannot be accounted for.

The other theoretical approach either employs analogy between a
strongly interacting 2D electron system and He$^3$ \cite{spivak} or
applies Gutzwiller's variational method \cite{br} to Si MOSFETs
\cite{dol}. It predicts that near the crystallization point, the
renormalization of $m$ is dominant compared to that of $g$ and that
the effective mass diverges at the transition. Although the sharp
increase of the mass is in agreement with our findings, it is the
expected dependence of $m$ on the degree of spin polarization that is
not confirmed by our data: the model of Ref.~\cite{spivak} predicts
that the effective mass should increase with increasing spin
polarization, whereas the prediction of the other model \cite{dol} is
the opposite.
%
%

Thus, the existing theories fail to explain our finding that in a
dilute 2D electron system the effective mass is strongly enhanced and
does not depend on the degree of spin polarization. The fact that the
spin exchange is not responsible for the observed mass enhancement
reduces somewhat the chances for the occurrence of an intermediate
phase --- ferromagnetic Fermi liquid --- that precedes electron
crystallization. In principle, should the spin exchange be small, the
spin effects may still come into play closer to the onset of Wigner
crystallization where the Fermi energy may continue dropping as
caused by mass enhancement.

In summary, we have found that in a dilute 2D electron system in
silicon, the strongly enhanced effective mass is independent of the
degree of spin polarization. This shows that the mechanism underlying
the effective mass enhancement is not related to spin and exchange
effects.
%
%

We gratefully acknowledge discussions with I.~L. Aleiner, D. Heiman,
and B. Spivak. This work was supported by NSF grant DMR-9988283, the
Sloan Foundation, the RFBR, the Russian Ministry of Sciences, and the
Programme ``The State Support of Leading Scientific Schools''.





\begin{thebibliography}{apssamp}
\bibitem{rem2} In principle, the exchange effects can also originate
from the isospin degree of freedom in bivalley (100)-Si MOSFETs. The
valley origin of the strongly enhanced effective mass is not very
likely, as inferred from a similar increase of the ratio of the spin
and the cyclotron splittings at low $n_s$ in the 2D electron system
in GaAs \cite{zhu}.
\end{thebibliography}
\end{document}